# Strain induced conductance modulation in graphene grain boundary

S. Bala Kumar and Jing Guo[a]

Department of Electrical and Computer Engineering, University of Florida, Gainesville, FL, 32611

[*]Corresponding author. E-mail: a) guoj@ufl.edu

Abstract

Grain boundaries (GBs) are ubiquitous in polycrystalline graphene materials obtained by various growth methods. It has been shown previously that considerable electrical transport gap can be opened by grain boundaries. On the other hand, polycrystalline graphene with GBs is an atomically thin membrane that can sustain extraordinary amount of strain. Here, by using atomistic quantum transport numerical simulations, we examine modulation of electrical transport properties of graphene GBs. The results indicate the modulation of transport gap and electrical conductance strongly depends on the topological structure of the GB. The transport gap of certain GBs can be significantly widened by strain, which is useful for improving the on-off ratio in potential transistor applications and for applications as monolayer strain sensors.

Graphene is a 2D material, with many interesting properties[1-3]. Similar to many other macroscopic materials, graphene exists in polycrystalline structures. The polycrystalline graphene samples haves been experimentally obtained using various techniques[4-6]. In a typical polycrystalline graphene, two 2D-graphene domains are separated by a 1D topological defect called grain boundary (GB), an intrinsic topological defect of the polycrystalline material[7]. In graphene, this GB is theoretically predicted to have interesting electronic[8-11], magnetic[12], chemical[13], mechanical[14-16] and thermal[17] properties.

The graphene GB appears to be a well ordered periodic structure, with periodicity of 1-5nm. This periodicity leads to the momentum conservation, which plays a significant role in the charge transmission across the GB[10]. Depending on the relative crystallographic orientations of the graphene domains, the GB can be either highly transparent or perfectly reflective for charge carrier over a large energy range. In other words, the transport gap in graphene could be modulated by proper engineering of the GB. This would enable the fabrication of graphene electronics, in which charge current can be controlled without the introduction of bulk bandgap in graphene.

Graphene, being one of the strongest materials, can withstand large strain[18]. The application of strain modifies the electronic structure of graphene. For example, the dirac cone is deformed and displaced under an uniaxial strain[18,19]. However, the transport gap in graphene remains zero, without being affected by the applied strain[18,19]. While the defects in graphene generally decrease the material's strength and thus decrease its ability to withstand large strain, it has been theoretically shown that graphene GBs, especially those with high density of defects, are as strong as pristine graphene.[15] In this letter, we investigate the effect of uniaxial strain on the transport properties across the graphene GB.

Here, we considered three different types of GB, i.e. symmetric-metallic, asymmetric-metallic, and semiconducting. Using atomistic quantum transport numerical simulations, we study the change in transport gap and conductance in these GBs due to the applied uniaxial strain. The symmetric GB is insensitive to strain while, a metal-to-semiconductor (semiconductor-to-metal) transition is obtained in the asymmetric-metallic (semiconducting) GB. We also develop an analytical model to approximate the transport gap as a function of strain for any generic GB structure. This would enable the usage of GB as a very sensitive strain sensor.

Referring to Fig. 1(a), the GB in graphene is a 1D interface between two domains of graphene, i.e. left-graphene-domain (LGD) and right-graphene-domain (RGD). The LGD and RGD have different crystallographic orientations. The structure of the GB depends on the relative orientation of the LGD and RGD[10]. GB consists of elementary topological defect, 5-pentagons and 7-heptagons. These elementary defects appears either individually, or as pentagon-heptagon pairs, or as series of such pairs, to fit the LGD-RGD misorientation. Here, we model the GB as a periodic array of defects[7,20], with the periodicity defined by exact/close matching of the translational vector, $\vec{v}_L = (n_L, m_L)$ and $\vec{v}_R = (n_R, m_R)$ corresponding to the LGD and the RGD, respectively. In Fig. 1(a), we shows the LGD, rotated such that the translational vector, $\vec{v} = (n, m)$. This results in a rotational angle, $\theta = \frac{\pi}{6} - \cos^{-1}\left(\frac{2m^2 + mn}{2m\sqrt{m^2 + mn + n^2}}\right)$ and the length of the repeat unit, $W = |\vec{v}| = a_0\sqrt{3}\sqrt{m^2 + mn + n^2}$. If there is a small mismatch between the length of the LGD repeat unit, $W_L$ and the RGD repeat unit, $W_R$, then we define the $W=(W_L+W_R)/2$.

At zero strain, the nearest neighbor bonding lengths, $r_i=a_0=0.142$nm and hoping parameter, $t_i=t_0=-2.66$eV, where i=1,2, ad 3 as labeled in the Fig. 1(a). When a uniaxial strain of $\delta$ is applied along the x direction, the graphene is deformed in the x and y direction from $x_0$ and $y_0$ to $x=x_0 p_x$ and $y=y_0 p_y$, where $p_x=1+\delta$ and $p_y=1-\eta\delta$. The Poisson ratio of the graphene, $\eta=0.15$.[21] As a result, the $r_i$ and thus the, $t_i$ are modified as follows:

$$r_1^2 = a_0^2 \cos^2(\theta+\pi/6) p_x^2 + a_0^2 \sin^2(\theta+\pi/6) p_y^2 \tag{1a}$$

$$r_2^2 = a_0^2 \cos^2(\theta-\pi/6) p_x^2 + a_0^2 \sin^2(\theta-\pi/6) p_y^2 \tag{1b}$$

$$r_3^2 = a_0^2 \cos^2(\theta+\pi/2) p_x^2 + a_0^2 \sin^2(\theta+\pi/2) p_y^2 \tag{1c}$$

, and the corresponding hopping parameter, $t_i=t_0(a_0/r_i)^2$.[22,23] We used a constant hoping parameter of $t_0$ across the GB. Note that, even under zero applied strain, in the vicinity of the GB, the actual hoping parameter would be different from $t_0$, due to the local strain caused by the atomic dislocation in the GB. However, while the quantitative value of the transmission across the GB is determined by the local strain, the qualitative behavior as well as the transport gap is not affected by the local strain.[10]

The electronic transport across the GB has been previously carried out using ab-initio calculations based on density-function-theory[16]. The results of the ab-initio calculations show a good agreement with the results obtain using π-orbital tight binding (π-TB) model. This indicates that the transport properties across the interface are mainly governed by the covalent bonds at the interface, and thus the π-TB is sufficient to describe the electronic transport across the GB. Therefore, here we investigate the electronic properties within the graphene domains, and across the GBs, using π-TB model. The real space π-TB Hamiltonian[24,25] of a graphene is $H = \sum_{\alpha,\beta} t_{\alpha,\beta} a_\alpha^+ a_\beta$ where $t_{\alpha,\beta}$ is the nearest neighbor hopping energy between atoms α and β. The electron transport behaviors are studied using the non-equilibrium Green's function (NEGF) formalism[26]. In the NEGF formalism, the electron transmission across the graphene, at Fermi energy, E is given by:

$$T(E,k) = Tr[\Gamma_L(E,k)G(E,k)\Gamma_R(E,k)G(E,k)^\dagger], \qquad (2)$$

where $G(E,k) = [EI - H(k) - \Sigma_L(E,k) - \Sigma_R(E,k)]^{-1}$ is the retarded Green's function of the GNR channel, $\Sigma_{L,R}(E,k)$ is the self energy coupling to the ideal graphene leads [see Supplementary Information], $\Gamma_{L,R}(E,k) = i[\Sigma_{L,R}(E,k) - \Sigma_{L,R}^+(E,k)]$, and k is the transverse wave vector along the GB direction. The subscript L(R) refers to the left (right) leads. As an example, the T(E,k) map of a graphene domain with $\vec{v} = (1,2)$ is shown in Fig. 1(b). At a finite temperature, τ the conductance across the graphene, at Fermi energy, E is

$$g(E,k) = g_0 \int_{-\infty}^{+\infty} T(E,k) \frac{1}{4k_B\tau} \text{sech}^2\left(\frac{\varepsilon - E}{2k_B\tau}\right) d\varepsilon, \qquad (3)$$

where $g_0 = e^2/h$ and $k_B$ is the Boltzmann constant.

Referring to Fig. 1(a), the vector, $\vec{v} = (n,m)$ can be divided into two cases, i.e. |n-m|=3p and |n-m|≠3p, where p is an integer. A GB is metallic when both the $\vec{v}_L$ and $\vec{v}_R$ corresponds to the same case, i.e. |n$_{L/R}$-m$_{L/R}$|=3p or |n$_{R/L}$-m$_{R/L}$|≠3p. A metallic GB is symmetric (asymmetric) when $m_{L/R} = n_{R/L}$ ($m_{L/R} \neq n_{R/L}$). On the other hand, a GB is semiconducting, when both the $\vec{v}_L$

and $\vec{v}_R$ corresponds to a different case, i.e. $|n_{L(R)}-m_{L(R)}|=(\neq)3p$ or $|n_{L(R)}-m_{L(R)}|\neq(=)3p$. Note that semiconducting GBs are always asymmetric, i.e. $m_{L/R} \neq n_{R/L}$. Here, we study the effect of strain in these three different types of GBs, i.e. (1) GB-I, a symmetric-metallic GB, with $\vec{v}_L=(2,1)$ and $\vec{v}_R=(1,2)$; (2) GB-II, a semiconducting GB, with $\vec{v}_L=(5,0)$ and $\vec{v}_R=(3,3)$; and (3) GB-III, asymmetric-metallic GB, with $\vec{v}_L=(5,3)$ and $\vec{v}_R=(7,0)$. The actual atomic structure of GB-I, GB-II, and GB-III is obtained from ref. 10.

First, we investigate the effect of strain on the charge transport across the GB. We computed the T(E,k) map for the GB structures, using the numerical π-TB approach. These results are plotted in Fig 2. GB-I and GB-III is metallic while GB-II is semiconducting when the δ=0. When a tensile strain is applied the transport gap increases for the GB-II and GB-III. Under compressive strain, the transport gap decreases (increases) for GB-II (GB-III), while the gap in GB-I is insensitive to the applied strain. Metal-to-semiconductor (semiconductor-to-metal) transition is observed in the GB-III (GB-II). It is worth noting that, while the ideal graphene domains exhibit the property of electron-hole symmetry [Fig. 1(b)], the transport across the GB does not respect the electron-hole symmetry. This is because the bipartite lattice of ideal graphene is locally broken at the GB, caused by the odd-membered rings.

To further understand the strain induced modulation of transport gap across GB, we plot the individual T(E,k) map, for the LGD and RGD at different applied strains, and investigate the effect of strain on these maps. Note that the LGD and RGD are rotated relative to the direction of the applied strain. We use a constant η, independent of the rotational angle. This leads to an equal deformation in the y-direction for both the LGD and RGD, when a constant strain is applied in the x-direction. One example of the simulated result, i.e. the LGD of GB-I, is shown in

Fig. 1(b). From this plot, we extract the values of $k_L$. We plotted a similar T(E,k) map for the RGD and extract the value of $k_R$. These $k_L$ and $k_R$ values as a function of strain are marked as "+" in the Fig. 3(b). Similarly, we extracted the $k_{L(R)}$ for GB-II and GB-III and plotted them in Fig. 3(c) and Fig. 3(d), respectively.

Under ballistic transmission E and k are conserved. Therefore, when the LGD and the RGD are connected by the GB, the transmission is possible only in the region where the T(E,k) of the LGD and RGD overlaps. In the Fig. 3(a), this is schematically indicated by the "darker" shade. As a result, a transport gap, Egap is obtained at $k=k_{gap}$, when $k_L \neq k_R$. The $E_{gap}$ and $k_{gap}$ can be approximated as follows

$$k_{gap}=(k_L+k_R)/2 \tag{4}$$

$$E_{gap} = \hbar v_F \left| k_L - k_R \right| \approx \frac{2.68 \left| K_L - K_R \right|}{\sqrt{m^2 + mn + n^2}} (eV) \tag{5}$$

, where the reduced plank constant, $\hbar = 1.054 \times 10^{-34} m^2 kg/s$, and fermi velocity $v_F = 10^6 ms^{-1}$, and $K_{L(R)}=Wk_{L(R)}$ is a dimensionless quantity obtained when W and $k_{L(R)}$ are in the units of m and m$^{-1}$, respectively. Eqs. 4 and 5, are valid when $v_F$ is independent to the applied strain. Even though, our numerical simulations show that, in some cases $v_F$ varies with the strain, these variations are too small, such that the Eqs. (4,5) remain a good approximation. Without the application of strain, for a semiconducting GB, the $k_L$-$k_R$=2π/3W and thus the $E_{gap} \approx 5.61/\sqrt{m^2 + mn + n^2}$ $(eV)$.

In general, graphene domains with different orientations show different response to the strain. GB-I is a symmetric-metallic- GB, with $\vec{v}_L = (2,1)$ and $\vec{v}_R = (1,2)$. Both left and right graphene domains has $|n-m| \neq 3p$, and thus the $k_L = k_R = 2\pi/3W$ and $E_{gap} = 0$, when there is no applied strain [refer Fig. 3(b)]. When a finite strain is applied, the $k_L$ and $k_R$ points shift to a different value. However, due to symetricity, the shift in $k_L$ and $k_R$ are equal. Therefore, the transport gap does not change, rather remains zero as the strain is varied.

On the other hand, GB-II is a semiconducting GB with $\vec{v}_L = (5,0)$ and $\vec{v}_R = (3,3)$. The LGD, has $|n-m| \neq 3p$ and the $k_L = 0$ under zero strain, while the RGD has $|n-m| = 3p$, and the $k_R = 2\pi/3W$. Thus when $\delta = 0$, there is a finite gap of $E_{gap} \approx 1eV$, at $k_{gap} = \pi/3W$. Referring to Fig. 3(c), when strain is applied, the dirac point of the LGD remains at $k_L = 0$. When a tensile (compressive) strain is applied, the $k_R$ moves to a higher (lower) value. As a result the transport gap is increased (decreased) with increasing tensile (compressive) strain. As shown in Fig. 3(f), the transport gap varies about 600%, i.e. from 0.25 to 1.5eV, by varying the strain from -20% to 20%.

GB-III is an asymmetric-metallic GB, with $\vec{v}_L = (5,3)$ and $\vec{v}_R = (7,0)$. Therefore, both the RGD and the LGD have dirac points at $k_L = k_R = 2\pi/3W$. Hence, this structure is metallic, with zero bandgap, when $\delta = 0$. As shown in Fig. 3(d), when a tensile strain is applied the $k_L$ decreases rapidly, while the $k_R$ does not change much. Thus the transport gap increases. When a compressive strain is applied, the $k_R$ increases to the maximum value of $\pi/W$ at $\delta = -0.12$. At this point, both the two valleys of the graphene meets at the $k = \pi/W$. Further increase in the compressive strain causes the other valley originating from k<0, to move closer to $k = 0^+$, resulting in decrease in $k_L$. Thus, the transport gap increases to a maximum value at strain $\delta = -12\%$, and then decreases, as we increase the compressive strain.

Note that the results is obtained by investigating the effect of strain on the individual LGD and RGD [refer Fig. 3(e-g)] is in very close agreement with the result simulation results across the GB [refer Fig. 2]. To get a deeper insight of the results in Fig. 3, we further derive an analytical expression for k as a function of the θ and δ. The Hamiltonian of the graphene is given by, $H = \begin{bmatrix} 0 & \lambda \\ \lambda^* & 0 \end{bmatrix}$, where

$$\lambda = t_3 + t_1 \exp\frac{ia_0\sqrt{3}(\sqrt{3}k_y' + k_x')}{2} + t_2 \exp\frac{ia_0\sqrt{3}(\sqrt{3}k_y' - k_x')}{2} \qquad (6)$$

The $k_x'$ and $k_y'$ are the longitudinal and transversal wave vector along the x' and y' direction. By solving for λ=0, the $k_{x(y)}'$ value at the dirac point, i.e. energy, E=0 is obtained as follows

$$k_x' = \frac{1}{\sqrt{3}a_0}\left(\cos^{-1}\frac{\alpha^2 - \beta^2 - 1}{2\beta} + \cos^{-1}\frac{\beta^2 - \alpha^2 - 1}{2\alpha}\right) \qquad (7a)$$

$$k_y' = \frac{1}{3a_0}\left(\cos^{-1}\frac{\beta^2 - \alpha^2 - 1}{2\alpha} - \cos^{-1}\frac{\alpha^2 - \beta^2 - 1}{2\beta}\right) \qquad (7b)$$

Where $\alpha = \frac{t_1}{t_3} = \frac{p_y^2 \sin^2(\theta + \pi/2) + p_x^2 \cos^2(\theta + \pi/2)}{p_y^2 \sin^2(\theta + \pi/6) + p_x^2 \cos^2(\theta + \pi/6)}$ and

$\beta = \frac{t_2}{t_3} = \frac{p_y^2 \sin^2(\theta + \pi/2) + p_x^2 \cos^2(\theta + \pi/2)}{p_y^2 \sin^2(\theta - \pi/6) + p_x^2 \cos^2(\theta - \pi/6)}$. The transversal wave vector along the GB direction, $k_y = k_y'\cos\theta - k_x'\sin\theta$. $k_y$ is the transversal wavevector in the 2D-Brillouin zone of the graphene. The equivalent transversal wavevector in the 1D-mini Brillouin zone of the periodic GB, $k \in [-\pi/W; \pi/W]$, is obtained by zone-folding method. Note that k=0 (k=2π/3W) if |n-m|=3p (|n-m|≠3p), consistent with the numerical simulations. The result of the

analytically obtain $k_L$ and $k_R$ values are plotted using the solid line in Fig 3(b-d). The solid curves matches perfectly with the "+" marks, indicating the $k_L(k_R)$ values calculated analytically matches perfectly with the values extracted from the numerically simulated T(E,k) map.

Finally, we study the conductance across the GB when different gate voltages are applied. Gate voltage would change the sheet charge density, and thus the Fermi energy. Fig 4 shows the variation of the conductances at different Fermi energy as a function of the varying strain. All simulations are done at the room temperature. Since, GB-I is always metallic, the conductance is insensitive to the applied strain. The transport gap in GB-II and GB-III, changes with strain, and thus conductance is very sensitive to the applied strain. For example, for the GB-II, at $E_F$=500meV, the conductance remain almost zero when a tensile strain is applied, while it increases rapidly when a compressive strain is applied. A similar effect is also seen for GB-III, where at $E_F$=250meV, the conductance decreases from 1eV to 0eV when 12% compressive strain is applied. This indicates the possibility of using GBs as strain sensor.

In the above simulations, we have assumed a periodic GB structure. The presents of disorders may break this periodicity, and thus introduce some leakage current within the transport gap region. However, previous studies[10] have shown that moderate disorders in GBs, cause only a low leakage current, and hence maintaining very low conductance within the transport gap. Furthermore, GB structures observed in recent experiments are not periodic[28-30], even though periodic GBs[10,13,14,15,27] are energetically the most preferred configurations [31]. This is because, in these observations, the GB structures are not well equilibrated, but rather frozen in disorder.[28] Unlike the periodic GBs which can be as strong as pristine graphene as indicated by the theoretical calculations[15], the aperiodic GBs are much weaker and thus can only withstand smaller strain.[28,30] Therefore, a controlled GB engineering, e.g. a proper annealing technique to create a periodic structure, would be beneficial to exploit the GB. For the sake of completeness, we also have presented the effect of compressive strain on the transport across the GB. However, with present technology it is not practically to induce compressive strain on any 2D material.

In conclusion, we showed that the GB, a topological defect, can be utilizes to modulate the transport gap, and thus the conductance in graphene, by applying uniaxial strain. It is found that the transport gap modulation under strain is sensitive to the degree of asymmetry of the GBs. While the symmetric GB remains metallic in the presence of uniaxial strain, the transport gap of the asymmetric semiconducting GBs can be considerably increased in the presence of strain. The asymmetric metallic GBs undergo a metal-semiconductor transition in the presence of strain. This property is very useful in utilizing the GB as highly sensitive strain sensor.

This work is supported by ONR and NSF. We would like to thank Y. Lu for helpful discussions.

Supporting Information Available: Calculation of self-energy for the semi-infinite graphene contacts. This material is available free of charge via the Internet at http://pubs.acs.org.

**References**

1. A. K. Geim, and K. S. Novoselov, Nature Mater. 6, 183 (2007).
2. M. I. Katsnelson, Mater. Today 10, 20 (2007).
3. A. H. Castro Neto, F. Guinea, N. M. R. Peres, K. S. Novoselov and A. K. Geim, Rev. Mod. Phys. 81, 109 (2009).
4. T. R. Albrecht, H. A. Mizes, J. Nogami, S-i. Park, and C. F. Quate, Appl. Phys. Lett. 52, 362 (1988).
5. C. R. Clemmer, and T. P. Beebe Jr, Science 251, 640 (1991).
6. J. Lahiri, Y. Lin, P. Bozkurt, I. I. Oleynik and M. Batzill, Nature Nanotech. 5, 326 (2010).
7. A. P. Sutton, and R. W. Balluffi, Interfaces in Crystalline Materials (Clarendon Press, 1995).
8. J. Cervenka, and C. F. J. Flipse, Phys. Rev. B **79**, 195429 (2009)
9. N. M. R. Peres, F. Guinea, and A. H. Castro-Neto, Phys. Rev. B **73**, 125411 (2006)
10. O. V. Yazyev, and S. G. Louie, Nature Mater. **6**, 806–809(2010)
11. A. Mesaros, S. Papanikolaou, C. F. J. Flipse, D. Sadri, and J. Zaanen, Phys. Rev. B **82**, 205119 (2010)
12. J. Cervenka, M. I. Katsnelson, and C. F. J. Flipse, Nature Phys. **5**, 840–844 (2009)
13. S. Malola, H. Hakkinen, and P. Koskinen, Phys. Rev. B **81**, 165447 (2010)
14. Y. Liu, and B. I. Yakobson, Nano Lett. **10**, 2178–2183 (2010)
15. R. Grantab, V. B. Shenoy, and R. S. Ruoff, Science **330**, 946–948 (2010).
16. O. V. Yazyev, and S. G. Louie, Phys. Rev. B **81**, 195420 (2010).
17. A. Bagri, Sang-Pil Kim, R. S. Ruoff, and V. B. Shenoy Nano Lett. 11, 3917 (2011).
18. M. Huang, H. Yan, T. F. Heinz and J. Hone, Nano Lett. 10, 4074 (2010).
19. V. M. Pereira and A. H. Castro Neto Phys. Rev. Lett. 103, 046801 (2009)
20. W. T. Read, and W. Shockley, Phys. Rev. 78, 275 (1950).
21. K. N. Kudin, G. E. Scuseria and B. I. Yakobson, Phys. Rev. B 64 235406 (2001).
22. W. A. Harrison, Electronic Structure and the Properties of Solids: The Physics of the Chemical Bond (Freeman, San Francisco, 1990).
23. L. Yang and J. Han, Phys. Rev. Lett. 85, 154 (2000).
24. R. Saito, M. Fujita, G. Dresselhaus, and M. S. Dresselhaus, Appl. Phys. Lett. 60, 2204 (1992).
25. A. H. C. Neto, F. Guinea, N. M. R. Peres, K. S. Novoselov, and A. K. Geim, Rev. Mod. Phys. **81**, 109 (2009).
26. S. Datta, Quantum Transport: Atom to Transistor, Cambridge University Press, Cambridge, (2005).
27. Johan M. Carlsson, Luca M. Ghiringhelli, and Annalisa Fasolino, PHYSICAL REVIEW B 84, 165423 (2011).
28. P. Y. Huang, C. Ruiz-Vargas, A. v.d. Zande, W. S. Whitney, M. P. Levendorf, J. W. Kevek, S. Garg, J. S. Alden, C. J. Hustedt, Y. Zhu, J. Park, P. L. McEuen, D. A. Muller, *Nature* 469, 389 (2011)
29. K. Kim, Z. Lee, W. Regan, C. Kisielowski, M. F. Crommie, and A. Zettl, ACS Nano 5, 2142 (2011)
30. Carlos S. Ruiz-Vargas, Houlong L. Zhuang, Pinshane Y. Huang, Arend M. van der Zande, Shivank Garg, Paul L. McEuen, David A. Muller, Richard G. Hennig, and Jiwoong Park, Nano Lett. 11, 2259 (2011).
31. B. I. Yakobson and F. Ding ACS Nano 5, 1569 (2011).

**Figures**

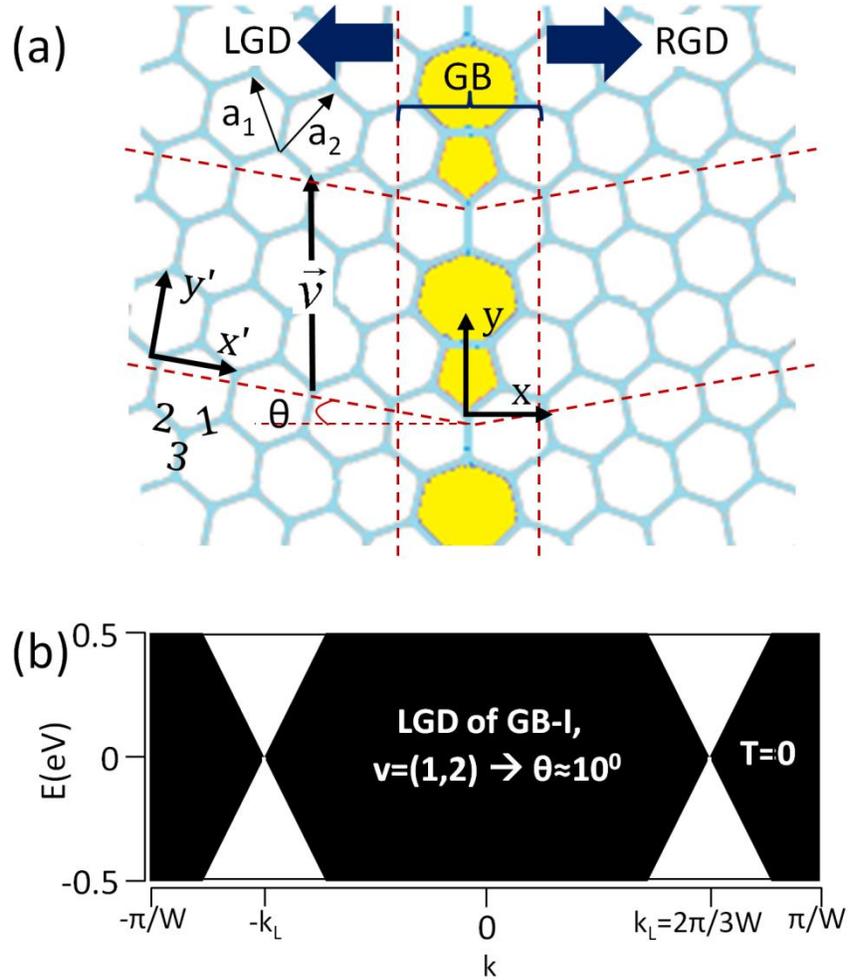

Figure 1 (a) Atomic structure of two graphene domains, i.e. left-graphene-domain (LGD) and right-graphene-domain (RGD) connected through a grain boundary (GB). The LGD of the grain boundary is rotated at the angle, $\theta$. The length of the repeat unit, $W = |\vec{v}|$. x(y)-direction is perpendicular (parallel) to the GB, while x'(y')-direction is rotated by $\theta$. The translational vector, $\vec{v} = m a_1 + n a_2 \equiv (m,n)$, where $a_1 = a_0(-\sqrt{3}x'+3y')/2$ and $a_1 = a_0(\sqrt{3}x'+3y')/2$. The labels: 1,2, and 3 refers to the three different nearest-neighbor directions for each atoms. (b) The T(E,k) map of LGD, when the $\vec{v} = (1,2)$. k is the wavevector in the y-direction, projected in the 1D mini Brillioun zone of the periodic GB. The dirac point is at $k=k_L$.

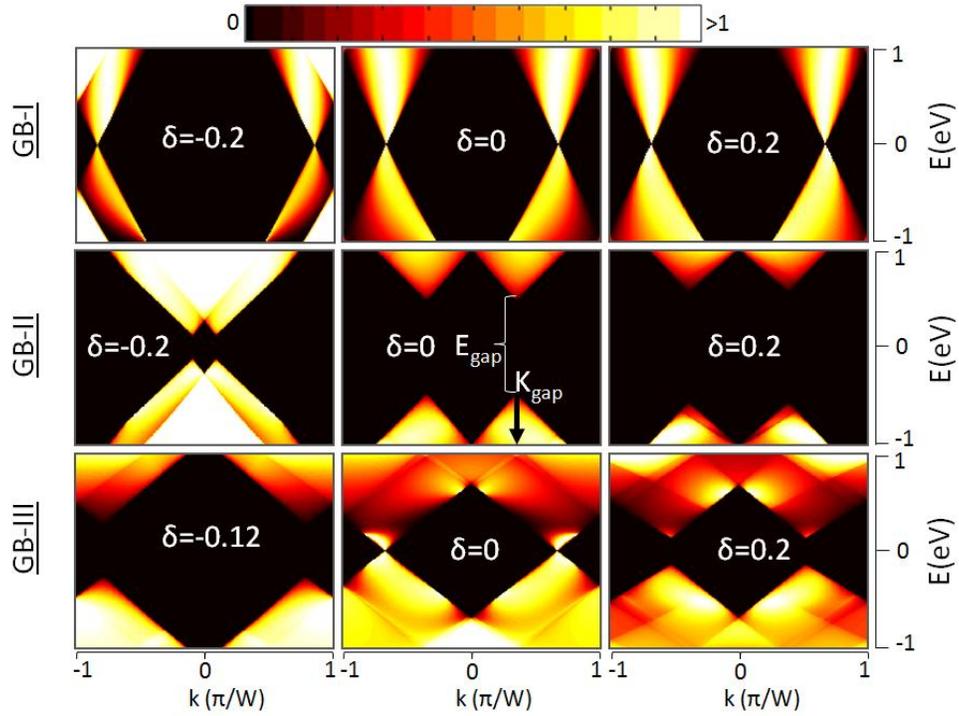

Figure 2 T(E,k) map for GB-I,II,III under compressive (δ<0), tensile (δ>0) and zero-strain (δ=0) condition. $E_{gap}$ and $k_{gap}$ are labeled. The $E_{gap}$ of GB-I is insensitive of strain. The $E_{gap}$ of GB-II, decreases (increases) with increasing compressive (tensile) strain. The $E_{gap}$ of GB-III, increases with increasing strain.

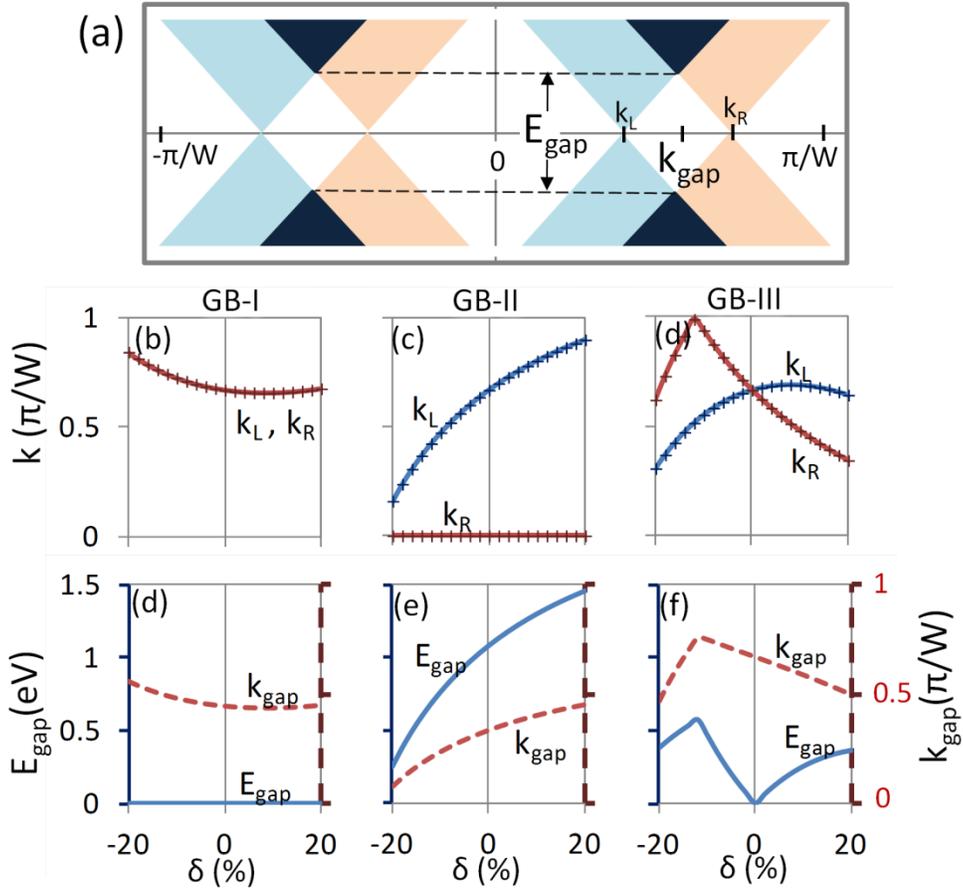

Figure 3 (a) The schematic illustration of the T(E,k) map of the LGD and RGD. $k_L$ and $k_R$, corresponds to the k value of the dirac point of the LGD and RGD, respectively. Electrons are transmitted only in the region where both the maps overlap. Thus when $k_L \neq k_R$, there is a finite transport gap, $E_{gap}$ at $k=k_{gap}$. (b-d) $k_L$ and $k_R$ values as a function of strain, for the LDG and RDG of different GBs. The solid lines are results computed analytically, while the marks "+" are extracted from T(E,k) map. (d-f) $E_{gap}$ and $k_{gap}$ values as a function of strain, for the LDG and RDG of different GBs. The repeat unit length, W=0.57nm, 1.10nm, and 1.50nm for GB-I, GB-II, and GB-III, respectively. The change of W due to strain is insignificant, i.e. $\Delta W = \pm 3\%$ for $\delta = \pm 20\%$.

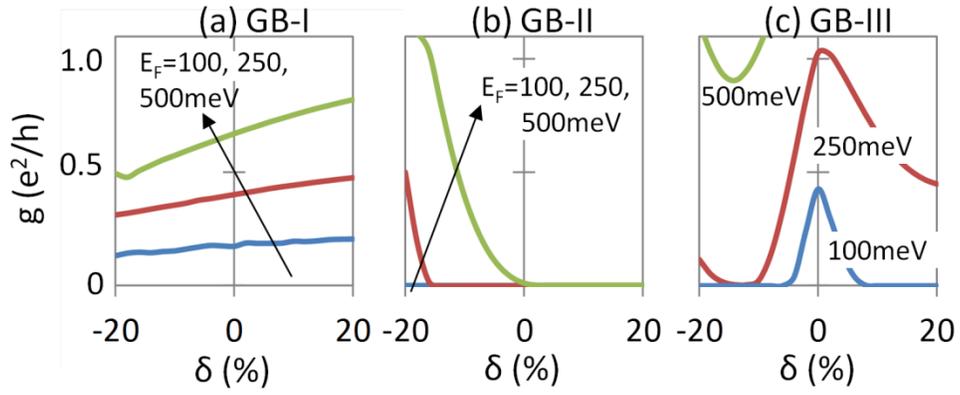

Figure 4 The application of gate voltage would change the charge sheet and thus the Fermi level, $E_F$. The change in the conductance across the GB, g as a function of strain at $E_F$=100,250,500meV for (a) GB-I, (b) GB-II, and (c) GB-III is shown. All these results are obtained at room temperature (300K).